\documentclass[aps,pra,twocolumn,showpacs,superscriptaddress,groupedaddress]{revtex4}
\usepackage{graphicx}  
\usepackage{dcolumn}   
\usepackage{bm}        
\usepackage{amssymb,amsmath}   

\begin{document}

\title{Detecting single photons is not always necessary to evidence interference of
photon probability amplitudes}

\author{Eric Lantz}
\email{eric.lantz@univ-fcomte.fr}
\author{Fabrice Devaux}
\affiliation{Institut FEMTO-ST,
D{\' e}partement d’Optique, UMR 6174 CNRS,
Universit{\' e} de Franche-Comt{\' e}, 15 B rue des Montboucons, 25030 Besan{\c c}on, France}
\author{Serge Massar}
\affiliation{Laboratoire d'Information Quantique CP224, Universit{\'e} libre de Bruxelles (ULB), Av. F. D. Roosevelt 50, B-1050 Bruxelles, Belgium}
\date{\today}

\begin{abstract}
Subtracting accidental coincidences is a common practice quantum optics experiments. For zero mean Gaussian states, such as squeezed vacuum, we show that if one removes accidental coincidences the measurement results are quantitatively the same, both for 
 photon coincidences at very low flux and for intensity covariances. Consequently, pure quantum effects at the photon level, like interference of photon wave functions or photon bunching, are reproduced in the correlation of fluctuations of macroscopic beams issued from spontaneous down conversion. This is true both in experiment if the detection resolution is smaller than the coherence cell (size of the mode), and in stochastic simulations based on sampling the Wigner function. 
 We  also discuss the limitations of this correspondence, such as Bell inequalities (for which one cannot substract accidental coincidences), highly multimode situations such as quantum imaging, and higher order correlations.
 
\end{abstract}

\maketitle

\section{Introduction}

Many iconic experiments in quantum optics, such as the Hong-Ou-Mandel (HOM) experiment \cite{Hong87}, demonstration of  Einstein-Podolsky-Rosen (EPR) position-momentum correlations\cite{Moreau14,Edgar12}, experimental tests of Bell inequalities \cite{Aspect82}, are based on correlations between detection of two photons. However, initial demonstrations of these experiments were done by subtracting accidental coincidences. Therefore these initial experiments measured the covariance of the detection rates. 

Later versions of these experiments were able to measure coincidences between single photon detections without subtraction of accidentals. For instance this was achieved for the  Hong-Ou-Mandel (HOM) experiment in \cite{Beugnon06, Somaschi16}, for demonstration of  Einstein-Podolsky-Rosen (EPR) position-momentum correlations in \cite{Courme23}, for experimental tests of Bell inequalities in \cite{Weihs98, Giustina15, Shalm15, Hensen15}.

One of the aims of the present paper is to clarify the interpretation of experiments in which covariance of detection rates is used. To  illustrate how these notions appears in quantum optics, consider the correlations between two beams impinging on two photodiodes $D_1$ and $D_2$ 
(as illustrated for instance in Fig. \ref{fig:setups} (a)).
To this end one can use the  mean of the product of the numbers $n_1$ and $n_2$ of photons  detected respectively on $D_1$ and $D_2$, with a delay $\tau$ between the detections $G^{(2)}_{12}(\tau)=<n_1 n_2>$, or its normalised version 
 $g^{(2)}_{12}(\tau)=G^{(2)}_{12}(\tau)\ /(<n_1><n_2>)$.
 Alternatively one can use the covariance $C_{12}(\tau)=<n_1n_2>-<n_1><n_2>$.
The first quantity is often used to characterise a single photon source. Indeed if 
  $D_1$ and $D_2$ are placed at the output of a balanced beam-splitter, then 
  $g^{(2)}_{12}(0)$ gives direct access to the purity of a single-photon source $1-g^{(2)}(0)$, where $g^{(2)}$ is the autocorrelation function of the beam before the beam-splitter \cite{Somaschi16}. (Indeed, it is easy to demonstrate that, in this detection scheme,  $g^{(2)}(\tau)= g^{(2)}_{12}(\tau)$). For a perfect single-photon source and at zero delay, we have $g^{(2)}(0)=0$, meaning that the detection of a photon on one photodiode prevents the simultaneous detection of a photon on the other photodiode, and consequently the covariance is negative. 
  On the other hand, the second quantity can be used to remove  accidental coincidences when a measurement would otherwise be affected by excessive noise. For instance, in the original HOM experiment \cite{Hong87}  performed with twin beams issued from Spontaneous Parametric Down Conversion (SPDC), see Fig. \ref{fig:setups} (b), the second quantity, i.e. the covariance, was measured, and  "suppression of coincidences" meant a zero covariance.
  
  The aim of photodetection is to measure the number operator $n$. Due to technological limitations this measurement is always imperfect and falls in two broad categories. Single photon detectors are generally of the on-off type: they are able to distinguish between the vacuum state $\vert 0 \rangle$ and states with one or more photons. When the average photon number $\langle n\rangle \ll 1$ is small, this is close to an ideal photon number measurement. These detectors are affected by several imperfections, such as dark counts and limited efficiency. On the other hand photodetectors, which are used at higher average photon number, produce a current proportional to the number of photons, but with added noise that prevents the the exact photon number to be resolved. Recently, photon number resolving detectors have been developed with the capability of resolving up to a dozen photons, see e.g. \cite{Miller03,Guo17,Cheng23}. (For simplicity, we will not consider such detectors in the present paper). One of the aims of the present work is to better understand the relation between experiments carried out in the low flux regime, in which on-off single photon detectors are used, and in the high power regime in which photodectors with a continuous output are used. Better understanding this relation will provide answers to the question raised in the title, namely,  do we really need single photon detectors to exhibit  interferences of individual photons? 
  
 For definiteness, in  the present work, we only consider the case, used in many experiments today \cite {Pan12}, in which photon pairs are produced via SPDC.
First of all we will show that when covariances are used, such experiments often have direct analogs in the high power regime, where twin beams are very strongly correlated, and in which the single photon detectors would be replaced experimentally by photodetectors that measure light intensities. In this regime the intensity correlations exhibit the same behavior as the  coincidence rate between single photon detections. This is true for Bell experiments using SPDC but we will see in section \ref{SubSecBell} that the classical reasoning leading to Bell inequalities involves products of intensities, not covariances. We summarize in Table \ref{tableau1} the status of subtraction of accidental coincidences in the different experiments considered in this paper.

In the present paper, we address some of the consequences of this equivalence between the low and the high power regime. First, the Hong-Ou-Mandel (HOM) experiment \cite{Hong87} can be considered as the first experiment evidencing the existence of a photon without destroying it: two indistinguishable photons interfere in a balanced beam-splitter and pursue together their path to one of the detectors, proving that photons do exist and are not a mere extrapolation of the quantification of the light-matter interaction \cite{Fabre13}. It is worth noting, and somewhat troubling, that this experiment can be performed at sufficiently high power with photodetectors that measure light intensities, not individual photons. Note however that even at high power the twin beams are highly quantum: even if the intensity in each beam fluctuates, the two beams experience exactly the same fluctuations, because of the photon pairs that constitute the beams. We aim to better understand the connection between the low and high power versions of this, and other, experiments.

A final important motivation for the present work concerns stochastic simulations of SPDC. This is a very useful tool for simulating quantum optics experiments, see e.g. \cite{Werner95,Werner97,Brambilla04,Devaux19, Devaux20}. 
Indeed, stochastic sampling of the Wigner function is by far the fastest method to simulate highly multimode quantum optics experiments, providing considerable speedup compared to computing the biphoton wavefunction.
However, as practitioners of this method know, using a (much) higher gain in simulations than in experiment
leads to qualitatively similar results, but with a large saving in computational time.
The present work helps undestand why one can use the high gain regime in simulations:
quantities like covariances will be similar in the low and high gain mode. We will illustrate this in Section \ref{SubSecStoch} with an example based on a multimode HOM experiment.

\begin{table}

\begin{tabular}{|l|c|c|}
\hline
  Experiment&\multicolumn{2}{c|}{Subtraction of} \\
  &\multicolumn{2}{c|}{accidental coincidences}\\ 
  \hline 
 & Very low flux & Higher flux \\ 
\hline 
 Characterization of  & Optional & Necessary \\ 
  twin photon sources&&\\
\hline 
HOM & Optional& Necessary \\ 
\hline 
Bell & Forbidden & Forbidden \\ 
\hline 
Purity of  & Forbidden &  Irrelevant \\ 
single photon sources&&\\\hline 
\end{tabular} 
\centering
\caption{Status of subtraction of accidental coincidences or subtraction of the product of mean intensities. }
\label{tableau1}
\end{table}

Finally we discuss limitations of the correspondence between low and high gain regimes.
We show that in experiments involving two photons, this correspondence  is not perfect in the multimode case because the gain experienced by the different modes may not be all equal, and may therefore not scale in the same way as one increases the pump power. 
This correspondence also breaks down in experiments involving more than two photons.
And Bell experiments of course require single photon detections. 
In Table \ref{tableau1} we list some iconic quantum optics experiments, and when the substraction of accidental coincidences is allowed, both in the low flux and high flux regimes.
 In the conclusion we discuss the implication of this correspondence for stochastic simulations of quantum optics experiments, as well as connections with hidden variable models.

\section{Stochastic field representation of symmetrised correlations}

To address the above mentioned correspondence between low and high power regimes, we use the Wigner representation as follows.

It was shown by Cahill and Glauber \cite{Cahill69} (equation (4.23)) that the expectation value of a symmetrically
ordered product of creation and annihilation operators $a^\dagger$ and $a$, can be always expressed as an integral in the
entire complex plane of the c-number $\alpha$ weighted by the Wigner function $W(\alpha)$:
\begin{equation}
\langle (a^\dagger)^n a^m \rangle_S = \frac{1}{\pi} \int_{\mathbb{C}} d^2 \ \alpha (\alpha^*)^n \alpha^m W(\alpha) 
\label{Eq:CahillGlauber}
\end{equation}

Furthermore it was shown in \cite{Reynaud92,Werner95,Werner97} that the Wigner function for a pump, signal and idler fields in a $\xi^{(2)}$ medium obey the classical equations of motion if the pump beam is undepleted, which is  a good approximation in the cases studied here. Hence, if the initial Wigner function is gaussian, then the Wigner function will stay gaussian. We will use this below.

When the initial Wigner function of the signal and idler fields is positive (which is the case if they are in the vacuum), then these results provide an efficient way to compute numerically symmetrised products of creation and annihilation operators in highly multimode situations. To this end one randomly samples the initial signal and idler fields  using as  probability distribution the initial Wigner function, and then propagates -through the nonlinear crystal, beam splitters, focusing optics, etc..-  these stochastic fields using the classical equations of motion. The average over the final distribution yields the desired expectation value. To obtain expectation of normal ordered operators a correction is required, for instance  subtraction of the constant 1/2 for the intensity, 1/4 for the variance (when expressed in units of photon number).

To illustrate the above we consider a SPDC experiment with two detectors at two distinct positions $D_1$ and $D_2$.  The symmetrised product of the corresponding field operators is given by the expectation of the stochastic classical fields 
\begin{eqnarray}
& \ &\langle E^\dagger_{D_1} E_{D_1}E^\dagger_{D_2} E_{D_2} \rangle_S\nonumber\\  
&\equiv &\frac{\langle (E^\dagger_{D_1} E_{D_1} +E_{D_1} E^\dagger_{D_1})  (E^\dagger_{D_2} E_{D_2} +E_{D_2} E^\dagger_{D_2})\rangle}{4}\nonumber\\  
&=& \langle E_{D_1}E_{D_1}^{*}E_{D_2}E_{D_2}^{*}\rangle
\end{eqnarray}
where on the left hand side we have the positive and negative frequency field operators and a quantum expectation value, and on the right the expection value of the classical stochastic fields. In the following we will generally work with the sochastic fields, and this will be obvious from the notation, as they get complex conjugated $E_{D_1}^{*}$ rather than Hermitian conjugated $E^\dagger_{D_1}$.

Since the Wigner function is Gaussian, the fields at ${D_1}$ and ${D_2}$ obey the Gaussian moment theorem  \cite {Mandel95, Isserli1918, Trajtenberg2020}:
\begin{eqnarray}
	\label{Gauss}
		\langle E_{D_1}E_{D_1}^{*}E_{D_2}E_{D_2}^{*}\rangle = & \langle E_{D_1}E_{D_1}^{*}\rangle \langle E_{D_2}E_{D_2}^{*}\rangle \nonumber \\
		+&\langle E_{D_1}E_{D_2}^{*}\rangle \langle E_{D_2}E_{D_1}^{*}\rangle \nonumber\\
		+& \langle E_{D_1}E_{D_2}\rangle\langle E_{D_1}^{*}E_{D_2}^{*}\rangle\ .
\end{eqnarray}

 Eq.(\ref{Gauss}) can be rewritten in terms of detected intensities  $I_{D_1}=E_{D_1}E_{D_1}^{*}$ and $I_{D_2}=E_{D_2}E_{D_2}^{*}$ as
 \begin{eqnarray}
	\label{Gauss2}
		\langle I_{D_1}I_{D_2}\rangle &= & \langle I_{D_1}\rangle \langle I_{D_2}\rangle 
		+
|\langle E_{D_1}E_{D_2}^{*}\rangle |^2+|\langle E_{D_1}E_{D_2}\rangle |^2		
\ .\nonumber
\end{eqnarray}
 Reorganising terms, we obtain an expression for the covariance:
\begin{eqnarray}
cov(I_{D_1},I_{D_2})&=&\langle I_{D_1}I_{D_2}\rangle -\langle I_{D_1}\rangle \langle I_{D_2}\rangle \label{covGauss2}\nonumber\\
&=&|\langle E_{D_1}E_{D_2}^{*}\rangle |^2+|\langle E_{D_1}E_{D_2}\rangle |^2\ . \label{covGauss}
\end{eqnarray}
Because of the subtraction of the product of the mean intensities (unlike in the $G_2$ coefficient), the covariance between the detected intensities can therefore be deduced from two mean products of two stochastic fields. In practice only one mean product remains: one of the two means vanishes because only phase differences make sense. This mean product of stochastic fields has a form similar to the product of creation operators in the biphoton wave function. Hence the behavior of the covariance of signal-idler intensities is the same as the correlation of the signal and idler photons in a pair. This is true in a vast number of quantum optics experiments using SPDC, both at very low fluxes, where photon coincidences are detected, as well as for intense twin beams where one measures intensities without photon resolution. It can be noted that the formalism of the biphoton concerns a single pair and consequently is valid in a regime of a very low flux where the probability of accidental coincidences is weak and can be neglected.  In such a regime, results are similar with and without subtraction of coincidences: as quoted in  Table 1, this subtraction is optional. On the other hand, at high flux this subtraction, or the subtraction of the product of mean intensities, is necessary to only  keep the same product of two fields as at low flux.

We note also that in the high intensity case one needs to use photodectors with resolution much smaller than the coherence cell (size of the mode), in the time as well as in the space domain. Otherwise, the fluctuations would be averaged out. On the other hand in the low intensity regime when single photon detectors are used, the time and space windows can be much larger than the coherence length, provided the probability of a single photon being detected in each window is much smaller than one. In this regime the subtraction of the accidental coincidences (the second term in Eq. (\ref{covGauss2})) is not compulsory. It improves the results, but the accidental coincidence term can be much smaller than the true coincidence term. The accidental coincidences in this regime could come from pairs in different modes or from dark counts. 

\section{Examples}

\subsection{Non-degenerate SPDC}

\begin{figure}[h!]
\begin{center}
\includegraphics[width=0.9\columnwidth]{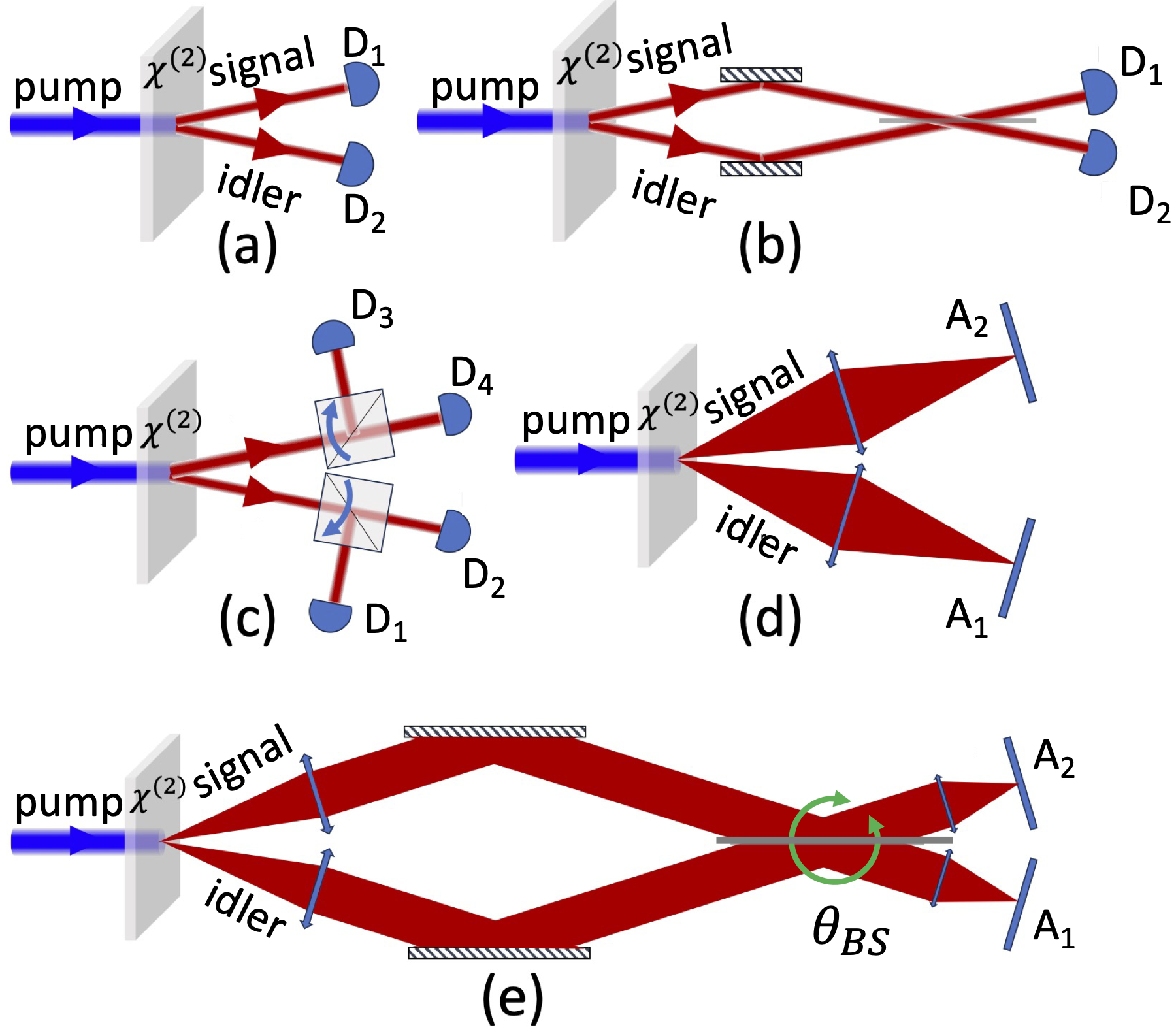} 
\end{center}
\caption{ Experimental setups considered in this work: a) non-degenerate Parametric Down Conversion (PDC); b) Hong-Ou-Mandel (HOM) experiment; c) Bell experiment; d) multimode non-degenerate PDC; e) HOM with multimode non-degenerate PDC. In panels (d) an (e), $A_1$ and $A_2$ are the imaging planes of cameras (possibly with single photon resolution). Pump beam is in blue; signal and idler photons produced by PDC are in red. In panel (e), $\theta_{BS}$ represents an angular shift of the beam-splitter, used in abcissa in Fig.\ref{dipnumeric}
}
\label{fig:setups}
\end{figure}

 We give now some examples of the use of (\ref{covGauss}) to establish well known results usually deduced from the biphoton wavefunction. These examples are illustrated in Fig. \ref{fig:setups}.
 
 We first consider the very simple case of  non-degenerate SPDC and a direct detection, at the output of the crystal, of the intensities $I_{s}$ and $I_{i}$ of respectively the signal and idler beams. At perfect phase-matching, the classical equations of parametric amplification gives, for a single pair of signal and idler fields $E_{s}$ and $E_{i}$,  after amplification in a crystal of length $L$:
 \begin{eqnarray}\label{Paramamplfier}
E_{s}(L)&=&C E_{s}(0)-i S  E_{i}^{*}(0) \nonumber\\
E_{i}(L)&=&C E_{i}(0)-i S  E_{s}^{*}(0)
\end{eqnarray}
where $C=\cosh(gL)$, $S=\sinh(gL)$,  with g the gain per length unit, proportional to the pump amplitude. The phases are defined with respect to the pump.   

 Using the fact that $E_{s}(0)$ and $E_{i}(0)$ are independent vacuum fields, with each a mean intensity  1/2, in units of photons per mode  \cite {Cahill69, Lantz21}, we obtain from (\ref{covGauss})
\begin{eqnarray}\label{covparam}
\left<E_s(L) E_s(L)\right>=\left< E_i(L) E_i(L)\right> = \left< E_s(L) E_i^*(L)\right> =0\nonumber
\end{eqnarray}
and
\begin{eqnarray}
 cov(I_s(L),I_i(L))&=&|\left< E_s(L) E_i(L)\right> |^2
 \nonumber\\
&=&C^2 S^2\left< E_{s}(0)E_s^*(0)+ E_i(0)E_i^*(0)\right>^2 
\nonumber\\
&=&C^2 S^2\ .
\end{eqnarray}

 We can also calculate the mean intensity and the variance:
 \begin{eqnarray}\label{varparam}
 \left <I_s(L)\right>&=&\left <E_s(L) E_s^*(L)\right> -1/2\nonumber\\
&=& C^2\left< E_s(0)E_s^*(0)\right>+S^2\left< E_i(0)E_i^*(0)\right>-1/2 
\nonumber\\
&=& S^2\nonumber
\end{eqnarray}
\begin{eqnarray}
 var(I_{s}(L))&=&|\left<E_s(L) E_s^* (L) \right>|^{2}-1/4\nonumber\\
&=&\left<C^2E_{s}(0)E_s^*(0)+S^2E_{i}(0)E_i^*(0)\right> ^2-1/4\nonumber\\ 
&=&C^2 S^2\ .
\end{eqnarray}
The subtraction of 1/2 for the intensity and  1/4 for the variance is necessary to pass from the symmetrized order to the normal order \cite{Cahill69, Lantz21}, while this correction is zero for covariances. 

We note that the variance of one of the twin beams is equal to the covariance between the beams. Thus, even at high flux, the beams are perfectly correlated.

The quantum efficiency $\eta$ of the detectors is easily taken into account by adding a fictitious beam-splitter that mixes the actual fields to vacuum:
\begin{eqnarray}\label{Qe}
 E_{D_1}&=&\sqrt{\eta} E_{s}+\sqrt{1-\eta}E_{v_1}\nonumber\\
 E_{D_2}&=&\sqrt{\eta} E_{i}+\sqrt{1-\eta}E_{v_2}
\end{eqnarray} 
where $E_{v_1}$ and  $E_{v_2}$ are two independent vacuum fields. As detailed in Appendix A, a straightforward calculation using (\ref{Qe},\ref{covGauss},\ref{Paramamplfier}) leads to:
\begin{eqnarray}\label{covQe}
 var(I_{D_1}) =var(I_{D_2})&=&\eta^2 S^4+\eta S^2\nonumber\\
  cov(I_{D_1},I_{D_2})&=&\eta^2 S^4+\eta^2 S^2\ .
\end{eqnarray}
The fluctuations of the classical intensity (first terms) remain perfectly correlated. This is to be contrasted to the shot-noise (second terms): random deletion of photons leads to detection of photons without twin.

\subsection{Hong Ou Mandel experiment} 
 
 We now pass to the HOM experiment \cite{Hong87}, see Fig 1 (b). As in the original experiment, we assume that the signal and idler photons are made indistinguishable by rotation of polarization and arrive on the two input ports s and i of a balanced and lossless  beam splitter, with output ports labeled 1 and 2. Conservation of energy imposes:  
 \begin{eqnarray}
 |t_{s_1}|^2+|r_{s_2}|^2&=&|t_{i_2}|^2+|r_{i_1}|^2=1;\nonumber\\
t_{s_1} r_{s_2}^{*}+r_{i_1}t_{i_2}^{*}&=&0
\label{BS}
 \end{eqnarray}
where $t_{k_l},(r_{k_l}), k=s,i, l=1,2$ are the transmission (reflection) coefficients from k to l, in amplitude.
We obtain: 
\begin{eqnarray}
 E_1&=&t_{s_1}E_s+r_{i_1}E_i;\nonumber\\
 E_2&=&r_{s_2}E_{s}+t_{i_2}E_{i}\nonumber\\
cov(I_{1},I_{2})&=&\left|\left<E_1 E_2^*\right>\right|^{2}+\left|\left<E_1 E_2\right>\right|^{2}\nonumber\\
&=&\left|(t_{s_1}t_{i_2}+r_{i_1}r_{s_2})<E_sE_i>\right|^{2}
\label{covBS}
\end{eqnarray} 
To establish  (\ref{covBS}), we have used (\ref{BS}),
 $\left<E_s E_s^*\right>=\left<E_i E_i^*\right>, \left<E_s E_s\right>=\left<E_i E_i\right>=\left<E_s E_i^*\right>=0$. If the beam-splitter is balanced, $t_{s_1}t_{i_2}+r_{i_1}r_{s_2}=0$, and we obtain as expected $cov(I_1,I_2)=0$. 
 
 We emphasize that this result holds both in the low gain (single photon) regime, and in the high gain (photodector) regime. 

 \subsection{Bell experiment} \label{SubSecBell}
 
 Our next example concerns the Bell state $\frac{1}{\sqrt{2}}\left(\left|H_1 V_2\right>+ \left|V_1 H_2\right>\right)$, where $1$ and $2$ design two distinct locations where the two photons of a pair are respectively detected, and $H$ and $V$ stand for horizontal and vertical polarisations, see Fig.1(c). SPDC is the most often used way to produce such entangled pairs; for instance
 Kwiatt et al. \cite {Kwiatt1995} have shown how to obtain such a state at the double intersection of the two cones of type-II SPDC. 
 
 If the signal (idler) is horizontally (vertically) polarized along $x$ $(y)$, then the fields $E_{1+}$ ($E_{2+})$ at location $1$ $(2)$ after passing through polarizing beam-splitters  oriented along $\theta_1$ ($\theta_2$) are written as:
 \begin{eqnarray}\label{Bell12}
 E_{1+}&=&E_{1x} \cos (\theta_1)+E_{1y} \sin (\theta_1)\nonumber\\
 E_{2+}&=&E_{2x} \cos (\theta_2)+E_{2y} \sin(\theta_2)\ .
\end{eqnarray} 
Using (\ref{covparam}), this  leads to a covariance between the respective intensities $I_{1+}$ and $I_{2+}$:
\begin{eqnarray}
 cov(I_{1+},I_{2+})&=&|E_{1+}E_{2+}|^2 \nonumber\\
 &=&|E_{1x}E_{2y} \cos(\theta_1) \sin(\theta_2)\nonumber\\
 & &+E_{1y}E_{2x} \sin(\theta_1) \cos(\theta_2)|^2\nonumber\\
 &=&C^2 S^2 \sin^2(\theta_1+\theta_2)\ .
 \label{covBell}
\end{eqnarray} 
By dividing by the variance given in (\ref{varparam}), we obtain the correlation coefficient
\begin{eqnarray}
 \rho = \frac{cov(I_{1+},I_{2+})
}{\sqrt{ var(I_{1+}) var(I_{2+})}} &=&  \sin^2(\theta_1+\theta_2)
 \label{corrBell}
\end{eqnarray} 
which has exactly the same form as the probability of detecting two photons in this configuration.
Thus, by replacing the product of the intensities by their correlations coefficient in the Clauser-Horne-Simony-Holt (CHSH) inequalities \cite {CHSH69}, we can retrieve, whatever the field intensity, the same value of the Bell parameter $\textit{B}$  as for a biphoton state, with a well known maximum of $2\sqrt{2}$.

However, the above does not provide a violation of the CHSH inequality because the positivity of intensities is one of the ingredients that is used to derive the CHSH inequality. For instance,  if we follow the reasoning presented in \cite {Walls08,Reid86}, there is a step which uses the inequality:
\begin{equation}
\left\vert\frac{I_+ - I_-}{I_+ + I_-}\right\vert< 1\ ,
\label{bellclass}
\end{equation}
where $I_+$ and  $I_-$ are the two output intensities of one of the polarizing beam-splitters. Using covariances is equivalent to replace $I_+$, ($I_-$) by $I_+ -\left<I_+\right>$, ($I_- -\left<I_-\right>$); but then these quantities are not always positive  and do not fulfill the condition (\ref{bellclass}).

It is nevertheless possible to use our approach to calculate the maximum gain $G$ that allows the Bell inequalities to be violated (giving only a threshold separating the regime of nonlocal pairs and the regime of correlated intensities, not a proof that this is the actual threshold). it is indeed easy to show from (\ref{Bell12}) and (\ref{Paramamplfier}) that, whatever the angles: 
\begin{equation}
G=\left<I_{1+}\right>=\left<I_{1-}\right>=\left<I_{2+}\right>=\left<I_{2-}\right>=S^2\ , 
\label{imoy}
\end{equation}
leading, using (\ref{covBell}), to:
\begin{eqnarray}
\left<I_{1+}I_{2+}\right>=\left<I_{1-}I_{2-}\right>&=&C^2 S^2 \sin^2(\theta_1+\theta_2)+S^4\nonumber\\
\left<I_{1+}I_{2-}\right>=\left<I_{1-}I_{2+}\right>&=&C^2 S^2 \cos^2(\theta_1+ \theta_2)+S^4
\label{i1i2}
\end{eqnarray}
Upon inserting this into the Bell expression $B$, one finds after some manipulations that
\begin{equation}
B(G) =  \frac{1+G}{1+3G} B(0)
\label{ThresholdBell}
\end{equation}
where $B(0)$ is the Bell parameter at vanishing gain, and $B(G)$ the Bell parameter at gain $G$. For violation of the CHSH inequality we need $B(G) > 2$, while $B(0)=2 \sqrt{2}$, 
leading to the threshold for violation of the CHSH inequality $\frac{1+G}{1+3G} < \frac{1}{\sqrt{2}}$. This was stated in a different but equivalent form in \cite{Reid86}, using the Heisengerg point of view, and in this form in \cite{Lantz21}, and then in \cite{Brewster21}. The approach developed here appears particularly simple. The details of the calculation are in Appendix \ref{appendix}

\subsection{Multimode case}

We now discuss how to take into account the  multimode character of the SPDC. 
We consider for definitiness spatial degrees of freedom, as illustrated in Fig.1(d), although the same reasoning would apply for temporal degrees of freedom. We consider two detector planes, and we
label the pixels of the detector array on the signal side with a subscript l and on the idler side with a subscript m. For a given crystal length $L$, the mean product of signal-idler fields used in the calculation of the covariance in (\ref{covparam}) can be expressed,  using the singular value decomposition \cite{Christ11}, as: 
\begin{equation}\label{svd}
\left< Es_l Ei_m\right>=\sum_k U_{lk}\lambda_k V_{mk}
\end{equation}
with $U$,$V$ unitary matrices and $0\leq\lambda_k$ the singular values. A gain $g_k$ can be defined for each mode $k$ as:
\begin{equation}
\lambda_k = \cosh(g_k)\sinh(g_k)=C_k S_k 
\end{equation}
The fields in each pair of Schmidt modes can be written as in (\ref{Paramamplfier}): 
\begin{eqnarray}\label{schmidtampli}
Es_{k}(L)&=&C_k Es_{k}(0)-i S_k  Ei_{k}^{*}(0) \nonumber\\
Ei_{k}(L)&=&C_k Ei_{k}(0)-i S_k  Es_{k}^{*}(0)\ .
\end{eqnarray}
The actual fields remain gaussian, since they result from the superposition of the gaussian Schmidt fields: 
\begin{eqnarray}\label{pixelfields}
Es_{l}&=&\sum_k U_{lk}Es_{k}\nonumber\\
Ei_{m}&=&\sum_k V_{mk}Ei_{k}\ .
\end{eqnarray} 
Because $U$ and $V$ are unitary, Eqs. (\ref {covparam},\ref {schmidtampli},\ref {pixelfields}) lead to (\ref{svd}). More generally, the unitarity of these matrices means that all the above monomode relations have a multimode equivalent, with a gain which is mode dependent. For example the intensity on a pixel on the signal side can be written in analoguous way to Eq. (\ref{varparam}):
\begin{eqnarray}\label{pixelintensity}
 Is_l&=&\left <Es_l Es_l^*\right> -1/2 \nonumber\\
&=&\left[\sum_k |U_{lk}|^2\left(C_k^2\left< Es_k(0)Es_k^*(0)\right>\right.\right. \nonumber\\
& &\left.\left.+S_k^2\left< Ei_k(0)Ei_k^*(0)\right>\right)\right]-1/2\nonumber\\
&=&\sum_k |U_{lk}|^2 S_k^2\ .
\end{eqnarray}

Analysing in detail the spatio-temporal variation of correlations is beyond the scope of this paper. The analysis in the temporal domain was already detailed in the original HOM paper \cite{Hong87}. For a numerical and experimental analysis of spatio-temporal effects in the HOM experiment, see \cite{Devaux19, Devaux20}. Here, we just would like to stress that the simple fact that the fields are gaussian means that Eq.(\ref{covGauss}) remains valid: for SPDC, the analysis of intensity covariances can be performed at high intensities and gives the same results as the analysis of photon coincidences at low flux. 

\begin{figure}
\begin{center}
  \includegraphics[width=1\linewidth]{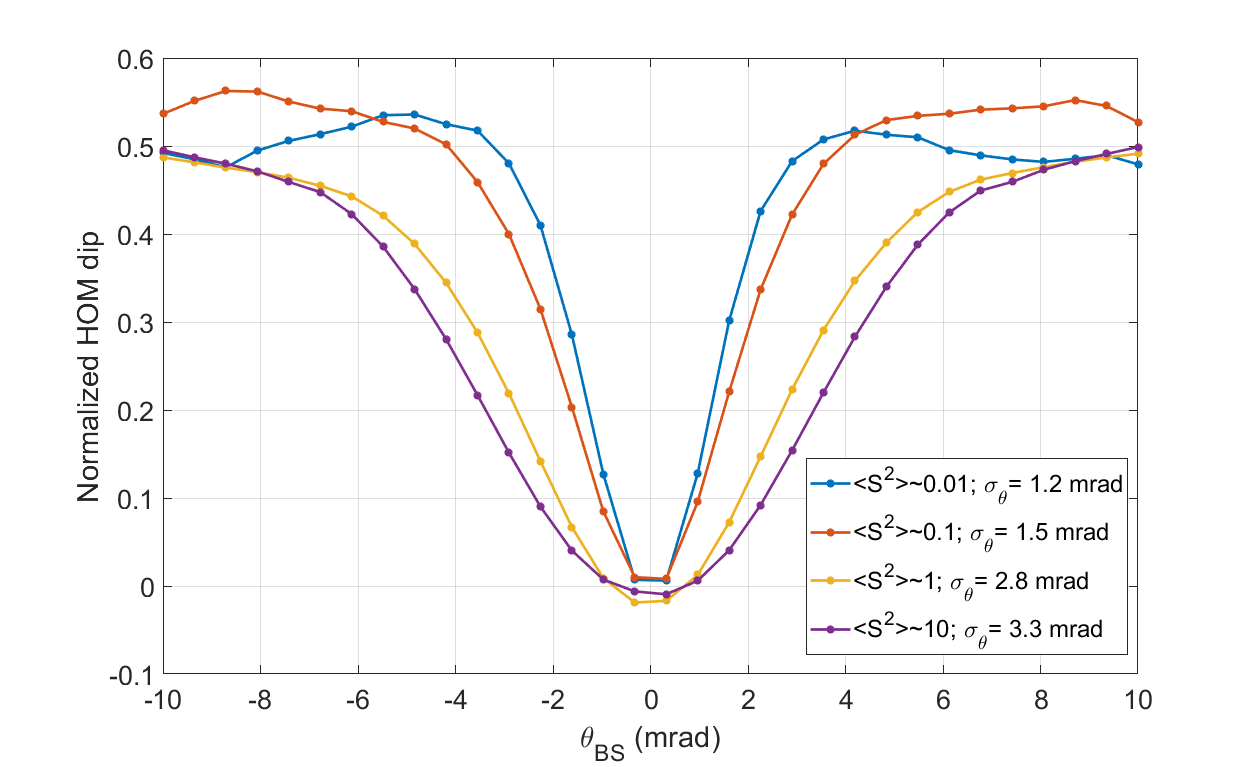}
 \caption{Influence of the gain $<S^2>$ in a multimode HOM experiment realised using the setup illustrated in Fig.\ref{fig:setups}(e), obtained using stochastic simulations. 
The gains in the different curves correspond to average number of photons per pixel of $0.01$, $0.1$, $1$, $10$. 
 In abscissa,  $\theta_{BS}$ is the rotation angle of the beam-splitter used to control momentum indistinguishability between the signal and idler beams. The horizontal angular shift of the reflected beams is twice $\theta_{BS}$. The results of numerical simulations using the stochastic model are averaged over 100 iterations. The crystal length is $L = 0.8mm$. In ordinates, the normalized amplitude of the HOM dip is obtained by dividing the correlation between the two output far-field images by the correlation between the two input far-field images. The correlation is obtained by dividing the covariance by the square root of the product of the variances, see Eq. (\ref{corrBell}). Note that as the gain changes by $4$ orders of magnitude, the width $\sigma_\theta$ changes by less than a factor of $3$.
(The width $\sigma_\theta$  given in inset is obtained  from gaussian fits to the numerical data). Note that the curves at low gain are more noisy (specially in the wings of the curves), contrary to the curves at high gain.}
  \label{dipnumeric}
\end{center}
\end{figure}

The main  difference with the single mode case is that, because the different modes have different gains, changing the power of the pump laser in a multimode SPDC experiment will not  simply multiply all intensities and correlation coefficients by a factor. 
Thus  a high gain allows the amplification of  modes that are not perfectly phase-matched. Hence, in a HOM experiment,(see Fig. 1 (e) for the spatial multimode case),a higher gain leads to a wider bandwidth in the spatial or temporal frequency domain.  This is illustrated in figure \ref{dipnumeric} by the simulation of  a spatial multimode HOM experiment with varying pump power. However the small dependence of the dip in  figure \ref{dipnumeric}  as a function of pump power means that one can use the simulations at high power, which are much faster, to describe with high precision the experiment in the low power, photon counting, regime.

\subsection{Multiparticle correlations} 

The above considerations have been made for correlations between two intensities. How does this generalise to correlations betweeen more than 2 intensities?
For definitiness, we consider  the four-fold correlation between two signal intensities, detected in two close locations  $s_1$ and $s_2$ separated by less than the size of the coherence cell, and two idler intensities in two close locations $i_1$ and $i_2$ \cite{Massar23}.\\ The four-fold covariance 
\begin{equation}
C_o=(I_{s_1}-\left<I_{s_1}\right>)(I_{s_2}-\left<I_{s_2}\right>)(I_{i_1}-\left<I_{i_1}\right>)(I_{i_2}-\left<I_{i_2}\right>)
\end{equation}
 has 9 terms. To see this, we reason as follows:\\
 - using the Gaussian moment theorem, we find that the fields $E_{s_1},E_{s_2},E_{i_1}^*,E_{i_2}^*$ give non zero terms only when associated to the fields $E_{s_1}^*,E_{s_2}^*,E_{i_1},E_{i_2}$, see (\ref{covparam}), resulting in $4!=24$ terms in $\left<I_{s_1}I_{s_2} I_{i_1}I_{i_2}\right>$.\\
-Hence $C_o$ has 24 terms from $\left<I_{s1}I_{s_2} I_{i1}I_{i2}\right>$, minus $4\times 6=24$ terms with a mean intensity times a three fold product like $\left<I_{s_1}\right>\left<I_{s_2} I_{i_1}I_{i_2}\right>$, plus $6\times 2=12$ terms with a product of two mean intensities, minus 4 terms with three mean intensities, plus one term product of the four mean intensities. The 15 removed terms include at least a mean intensity, and this is precisely the number of terms in $\left<I_{s_1}I_{s_2} I_{i_1}I_{i_2}\right>$ which include such a mean. 

Hence, $C_o$ includes only the 9 terms without a mean intensity:
\begin{eqnarray}
C_o=& \left<E_{s_1}E_{s_2}^*\right>\left<E_{s_2}E_{s_1}^*\right>\left<E_{i_1}^*E_{i_2}\right>\left<E_{i_2}^*E_{i_1}\right> \label{A1} \\
&+\left<E_{s_1}E_{s_2}^*\right>\left<E_{s_2}E_{i_1}\right>\left<E_{i_1}^* E_{i_2}\right>\left<E_{i_2}^*E_{s_1}^*\right>\label{A2}\\
&+\left<E_{s_1}E_{s_2}^*\right>\left<E_{s_2}E_{i_2}\right>\left<E_{i_1}^*E_{s_1}^*\right>\left<E_{i2}^* E_{i_1}\right>\label{A3}\\
&+\left<E_{s_1}E_{i_1}\right>\left<E_{s_2}E_{s_1}^*\right>\left<E_{i_1}^* E_{i_2}\right>\left<E_{i_2}^*E_{s_2}^*\right>\label{A4}\\
&+\left<E_{s_1}E_{i_1}\right>\left<E_{s_2}E_{i_2}\right>\left<E_{i_1}^* E_{s_1}^*\right>\left<E_{i_2}^*E_{s_2}^*\right>\label{A5}\\
&+\left<E_{s_1}E_{i_1}\right>\left<E_{s_2}E_{i_2}\right>\left<E_{i_1}^* E_{s_2}^*\right>\left<E_{i_2}^*E_{s_1}^*\right>\label{A6}\\
&+\left<E_{s_1}E_{i_2}\right>\left<E_{s_2}E_{s_1}^*\right>\left<E_{i_1}^*E_{i_2}\right>\left<E_{i_2}^* E_{s_2}^*\right>\label{A7}\\
&+\left<E_{s_1}E_{i_2}\right>\left<E_{s_2}E_{i_1}\right>\left<E_{i_1}^*E_{s_1}^*\right>\left<E_{i_2}^*E_{s_2}^*\right>\label{A8}\\
&+\left<E_{s_1}E_{i_2}\right>\left<E_{s_2}E_{i_1}\right>\left<E_{i_1}^*E_{s_2}^*\right>\left<E_{i_2}^*E_{s_1}^*\right>\ .\label{A9}
\end{eqnarray}
These terms scale differently as a function of the gain.

The term (\ref{A1}) is a term of incoherent bunching, proportional to $S^8$.\\
The terms \begin{eqnarray}
& &(\ref{A5})+(\ref{A6})+(\ref{A8})+(\ref{A9}) =\nonumber\\
& & \quad \quad \left|E_{s_1}E_{i_1}E_{s_2}E_{i_2}+E_{s_1}E_{i_2}E_{s_2}E_{i_1}\right|^2\ ,  
\end{eqnarray}
 proportional to $C^4S^4$, are  the only ones that remain at low gain. They can result in interferences in the four-photon coincidences \cite{Massar23}. \\
The four remaining terms, proportional to $C^2S^6$, have a less clear interpretation. They can be written as:
\begin{equation}
	\begin{aligned}
	& (\ref{A2})+(\ref{A3})+(\ref{A4})+(\ref{A7})=\{\left<E_{s_1}E_{s_2}^*\right>  \\
	&
	\times (\left<E_{s_2}E_{i_1}\right>\left<E_{i_1}^* E_{i_2}\right>\left<E_{i_2}^* E_{s_1}^*\right> \\
&\quad +\left<E_{s_2}E_{i_2}\right>\left<E_{i_1}^* E_{s_1}^*\right>\left<E_{i_2}^* E_{i_1}\right>) \}\\
 & \quad + \text{cc}\ .
\end{aligned}
\end{equation}

These relations can be extended to the multimode case, most easily using the singular value decomposition Eq. \ref{svd}.

\section{Discussion}

\subsection{Correspondence between low and high power experiments}

The first message conveyed here is that purely quantum effects, intimately linked to the particle character of photons, have their exact counterpart in the fluctuations of macroscopic twin beams. These macroscopic twin beams are not classical beams: they are formed by pairs and possess quantum properties \cite{Treps05}. For example, if the photon number of each beam strongly fluctuates with a thermal statistics, the fluctuations of both beams are strictly the same and the variance of the difference of photon numbers is exactly zero in an ideal experiment \cite {Agafonov10}. A practical illustration is the use in quantum imaging experiments  of  a variance of the difference of photon number smaller than the Poisson noise to prove the particle character of twin images \cite {Jedrkiewicz04, Blanchet08, Brida09, Moreau14}. Furthermore, with twin macroscopic beams, the visibility of interference in a HOM experiment is not limited to $0.5$, as it is the case for classical beams \cite {OuMandel89,Kim13}, but can go down to zero. 
Thus the HOM interference for a photon pair can be generalized to the interference of many photon pairs in a single mode, with covariance as the quantity that is used in both situations. Subtracting accidental coincidences is not only a useful procedure to eliminate the effect of independent pairs coming from other modes (or electronic noise), but also to take into account correlated pairs in a mode, obeying a Bose-Einstein (thermal) statistics. If the use of the subtraction of accidental coincidences to remove noise coming from other modes is quite obvious, the use of covariance even inside a single mode is much less intuitive, but is correct for Gaussian statistics. Bell inequalities are an exception: they use products of intensities, not their covariance.  Indeed, the violation of Bell inequalities describes the non local character of the correlation of two photons forming an unique pair.

\subsection{Stochastic simulations}\label{SubSecStoch}

The second message concerns stochastic simulations, which is a very useful tool for simulating quantum optics experiments, see e.g. \cite{Werner95,Werner97,Brambilla04}. 
Indeed, stochastic sampling of the Wigner function is by far the fastest method to simulate highly multimode quantum optics experiments, such as quantum imaging involving an image of $N$ by $N$ pixels.  Indeed the computation time will be proportional to $ N^2$. In comparison, the computation time of the biphoton wave function is at least proportional to $N^6$  \cite {Soro21}. 
For recent illustrations of application of this method, we refer to highly spatially multimode HOM experiments presented in \cite {Devaux19, Devaux20},

As practitioners of this method know, using a (much) higher gain in simulations than in experiment leads to qualitatively similar results, but with a large saving in computational time.
The reasons for the qualitative similarity of the results is explained by the present analysis: quantities like covariances will be similar in the low and high gain mode. The (small) differences are due for instance to phase matching conditions: a high gain allows the amplification of some modes that are not perfectly phase-matched \cite{Yariv89}. As illustration, in a HOM experiment, a higher gain leads to a slightly wider bandwidth in the spatial or temporal frequency domain.  This is illustrated in figure \ref{dipnumeric} for the spatial HOM experiment described in Figure \ref{fig:setups} (e). We see that, whatever the gain, the covariance falls close to zero in the center of the dip, validating the use in numerics of a higher gain than in the experiment.

The advantage in simulation of using a high gain is drastic: stochastic simulations with a low gain imply a huge number of repetitions of the simulation to obtain an acceptable signal-to-noise ratio (SNR). This is due to the fact that the signal has the level of the actual intensity without the input vacuum noise, while the noise includes this vacuum noise \cite {Blanchet10}. Thus for a small gain the SNR (defined as the ratio between the mean intensity and its standard deviation) will be equal to the average number of photons per mode (which in experimental situations will be $0.1$ or smaller).
On the other hand, with a gain of many photons per mode, the influence of the input vacuum noise on the SNR becomes negligible. Thus with a high gain and for one repetition, the SNR, is equal to one, because of the Bose-Einstein statistics of the intensity. Hence, for $R$ repetitions, the SNR is equal to $\sqrt{R}$. The total computational cost will thus be $R\times N^2$.

On the experimental side, using high gain to demonstrate quantum effects at the photon level is possible in principle but less evident than in simulations. The principal reason is that the time resolution of the detectors is often much greater than the duration of a SPDC mode, given  by the inverse of the spectral bandwidth of phase-matching, or of the added spectral filter if any. In the 1987 HOM experiment \cite{Hong87}, the time window for coincidences of twin photons was $7\: ns$, for a coherence time of about $100\: fs$. With such time scales, the fluctuations of more than $10^4$ independent modes would be averaged if using a high intensity beam. An obvious solution to work in the quasi-monomode regime is to use short pulses and narrow spectral filters with a time separation of the pulses smaller than the time window of the detectors. This was done in \cite {Mosset04, Jedrkiewicz04, Eisenberg04, Harder16}.  Eisenberg {\it et al.} \cite {Eisenberg04} analyzed two-photon coincidences with a small quantum efficiency and up to 50 photons per mode. However, their use of on-off detectors leads to a strong distortion of the statistics. For example, at high fluxes, the probability of coincidences tends to one and the covariance tends to zero. In \cite {Harder16}, photon number resolving detectors were employed. For images, Jedrkiewicz {\it et al.} demonstrated \cite {Jedrkiewicz04} the sub-shot noise character of the difference between signal-idler images issued from type II SPDC with 100 photons per mode. In \cite {Mosset04}, high intensity  images produced by type I SPDC where obtained and used to demonstrate the Bose-Einstein character of the statistics, and an image is reported in which 
 the twin character of the signal-idler fluctuations is clearly visible, but not quantitatively analyzed. 

\subsection{A hidden variable model?}

Stochastic simulations use the propagation of classical fields to reproduce the predictions of quantum experiments. We discuss here their connection with hidden variable models of quantum mechanics. Such
hidden variable models are interesting because they can in some cases provide an intuitive, classical, picture of the underlying quantum phenomena. 
Ideally  such a hidden variable model should have the following characteristics:
\begin{enumerate}
\item The hidden variable model reproduces the outcomes of one or several observables. That is, individual realisations of the hidden variable model predict individual outcomes of the observable, with the correct probability distribution being reproduced when averaged over the hidden variables.
\item
The hidden variable model is local, that is when describing multiparticle systems, one can assign hidden variables to each particle, and the evolution of these hidden variables depend only on the local environment of each particle. 
\end{enumerate}
Bell's theorem \cite{Bell87} proves that one cannot satisfy both requirements. Bohm's model \cite{Bohm52} shows that the first requirement can be met for position measurements of single particles. But Bohm's model, or extensions thereof,  cannot be extended to a local model of two or more entangled particles (since otherwise a contradiction with Bell's theorem would obtain).

The stochastic  simulations of quantum optics experiments satisfy the second requirement since the fields are propagated using the classical equations of motion, and hence are local, but do not satisfy the first requirement. To see this explicitly, consider the symmetrised number operator
$ (a^\dagger a + a a^\dagger)/2$. This operator has half integer eigenvalues $1/2, 3/2, ...$. The stochastic model will, at each repetition, yield a postive real value for the symmetrized number operator. The average will yield the correct expectation value, see Eq. (\ref{Eq:CahillGlauber}). But the individual runs cannot be used so simulate individual outcomes of the measurement (otherwise a contradiction with Bell's theorem would obtain). This can also be seen from the fact that individual runs can yield values less then $1/2$, corresponding to negative photon number, which would be unphysical.

 A fundamental difference remains however between the low and the high flux regime. 
 At high flux, a detection proportional to intensity is described by projection onto a positive Wigner function, meaning that one repetition of the experiment or of the corresponding stochastic simulation can be drawn from the same probability distribution \cite {Mari12, Rahimi16}. On the other hand, the on-off detectors used at low flux correspond to a projection on a one-photon state with a partially negative Wigner function. In this regime, as mentioned above,  there is no  correspondence between the experimental and simulated samples. For example, a sample in simulation can correspond to a negative intensity after the substraction required to obtain the normal ordered operator. Only the covariance, i.e. a mean over a large number of repetitions, is identical in simulations and experiments. 
  
  As a final remark, an alternative exists to simulate measurement outcomes at very low flux: the photon pairs can be considered as independent and the probability distribution can be directly inferred from the squared biphoton amplitude. 
  The simulation of  experimental samples, i.e. sampling from the classical probability distribution, appears particularly difficult in the intermediate situation, with a gain neither much lower nor much higher than one. Indeed this is a situation  similar to boson sampling, where it is expected that such sampling is computationally hard. For a study of quantum imaging experiments in this regime, see \cite{Massar23}.

\subsection{Summary} 

In the present work we have not presented any new quantum optics results. Rather we have clarified the connection between low flux and high flux quantum optics experiments, through the lens of stochastic field simulations.
For Gaussian states like SPDC, we have shown that the covariance  describes coincidences of photons at very low flux as well as correlations of intensity fluctuations at high flux. We have retrieved for covariances some well known results for coincidences, like the disappearance of coincidences in a HOM experiment. The computations are analogous to those based on the biphoton wave function, but valid for any number of photon pairs in a mode. We have only treated some simple cases, but the extension to more realistic or more complex situations,  can be readily carried out. The same results  can of course also be obtained by using  the Heisenberg representation, but the connection with the biphoton state is less evident. Our work also shows why high gain stochastic simulations of  an experimental set-up (which are computationaly efficient) will generally yield results close to those obtained in the experimental, low gain, regime. Finally the conceptual link to hidden variable models was discussed.

\appendix
\section{derivation of Eq.(\ref{covQe})} 
\renewcommand{\theequation}{A\arabic{equation}}
\setcounter{equation}{0}

We detail here the computation leading to the expressions (\ref{covQe}) of the variance and covariance for a non unity quantum efficiency $\eta$. As stated in the main text, the addition of a fictitious beam-splitter before the detectors leads to Eq. \ref{Qe}:
\begin{eqnarray}
 E_{D_1}&=&\sqrt{\eta} E_s(L)+\sqrt{1-\eta}E_{v_1}\nonumber\\
 E_{D_2}&=&\sqrt{\eta} E_i(L)+\sqrt{1-\eta}E_{v_2} ,\nonumber
\end{eqnarray} 
leading to a mean intensity on the photodiode $D_1$: 
\begin{eqnarray}\label{etaintensity}
 \left<I_{D_1}\right>&=& \left< E_{D_1}E_{D_1}^{*}\right>-1/2\nonumber\\
&=&\left<E_s(L) E_s^{*}(L)\right>+(1-\eta)\left< E_{v_1}  E_{v_1}^{*}\right> -1/2\nonumber\\
&=& \eta(C^2+S^2)/2+(1-\eta)/2-1/2 \nonumber\\
&=& \eta S^2 \ .
\end{eqnarray}
As expected, the intensity is simply multiplied by the quantum efficiency.\\
 The variance is computed in the same way:
\begin{eqnarray}\label{varQeAn}
var(I_{D_1})
&=&\left<(E_{D_1}E_{D_1}^{*})^2\right>-1/4\nonumber\\
&=&\eta^2((C^2+S^2)/2)^2+\eta(1-\eta)(C^2+S^2)/2\nonumber\\
& &+1/4(1-\eta)^2-1/4\nonumber\\
&=&\eta^2 S^4+\eta S^2 \ .
\end{eqnarray}
The computation of the covariance between the intensities in $D_1$ and $D_2$ is simpler, since the vacuums $v_1$ and $v_2$ are not correlated:
\begin{eqnarray}\label{covarQeAn}
cov(I_{D_1},I_{D_2})&=&|\left< (E_{D_1}E_{D_2})\right>|^2\nonumber\\
&=&\eta^2|\left< E_s(L) E_i(L)\right>| ^2\nonumber\\
&=&\eta^2 C^2 S^2=\eta^2 S^4+\eta^2 S^2 \ .
\end{eqnarray}

\section{Derivation of Eq. (\ref{ThresholdBell}). }\label{appendix}
\renewcommand{\theequation}{B\arabic{equation}}
\setcounter{equation}{0}
The coefficients used in the CHSH inequalities have the form:
 \begin{eqnarray}
E(\theta_1,\theta_2)&=&\frac{\left<I_{1+}I_{2+}+I_{1-}I_{2-}-I_{1+}I_{2-}-I_{1-}I_{2+}\right>}{\left<I_{1+}I_{2+}+I_{1-}I_{2-}+I_{1+}I_{2-}+I_{1-}I_{2+}\right>}\nonumber\\
&=&\frac{2 C^2 S^2 (\sin^2(\theta_1+\theta_2)-\cos^2(\theta_1+\theta_2))}
{2 C^2 S^2+4 S^4}\nonumber\\
&=&\frac{(1+G)(\sin^2(\theta_1+\theta_2)-\cos^2(\theta_1+\theta_2))}{1+3G}\ .
\label{highG}
\end{eqnarray}
Hence, for a non negligible gain $G$, the coefficient $\textit{B}$ is multiplied by $\frac{1+G}{1+3G}$, preventing any violation of the CHSH inequalities for $\frac{1+G}{1+3G} \leqslant \frac{1}{\sqrt{2}}$.


\begin{thebibliography}{100}
\bibitem{Hong87} C. K. Hong, Z. Y. Ou, and L. Mandel, \textit{Measurement of Subpicosecond Time Intervals between Two Photons by Interference}, Physical Review Letters {\bf 59}, 2044 (1987)


\bibitem{Moreau14} P. A. Moreau, F. Devaux, and E. Lantz, \textit{ Einstein-Podolsky-Rosen paradox in twin images}, Physical Review Letters {\bf 113},  160401 (2014)

\bibitem{Edgar12} M. P. Edgar {\it et al.},  \textit{Imaging high-dimensional spatial entanglement with a camera}, Nature Communications {\bf 3}, 984 (2012)

\bibitem{Aspect82} A. Aspect, J. Dalibard, and G. Roger, \textit{Experimental Test of Bell's Inequalities Using Time-Varying Analyzers} Phys. Rev. Lett. \textbf{49},1804 (1982)

\bibitem{Beugnon06}J. Beugnon {\it et al.}, \textit{Quantum interference between two single photons emitted by independently trapped atoms} Nature \textbf{440}, 779-782 (2006)

\bibitem{Somaschi16} N. Somaschi {\it et al.}, \textit{Near-optimal single-photon sources in the solid state}, Nature Photonics {\bf 10}, 340 (2016)


\bibitem{Courme23} B. Courme {\it et al.}, \textit{Quantifying high-dimensional spatial entanglement with a single-photon-sensitive time-stamping camera},
 Opt. Lett. \textbf{48}, 3439-3442 (2023) 

\bibitem{Weihs98} G. Weihs \textit{et al.},\textit{Violation of Bell’s Inequality under Strict Einstein Locality Conditions}
Phys. Rev. Letters \textbf{81}, 5039 (1998)

\bibitem{Giustina15}M. Giustina \textit{et al.,} \textit{Significant-Loophole-Free Test of Bell’s Theorem with Entangled Photons} Phys. Rev. Letters \textbf{115}, 250401 (2015)

\bibitem{Shalm15}L. K. Shalm \textit{et al.},\textit{Strong Loophole-Free Test of Local Realism} Phys. Rev. Letters \textbf{115}, 250402 (2015)

\bibitem{Hensen15} B.Hensen \textit{et al.,} \textit{Loophole-free Bell inequality violation using
electron spins separated by 1.3 kilometres}, Nature \textbf{526}, 682 (2015)


%
%

\bibitem{Miller03} A. J. Miller, S. W. Nam, J. M. Martinis, and A. V. Sergienko, \textit{ Demonstration of a low-noise near-infrared photon counter with multiphoton discrimination}, Appl. Phys. Lett. 83, 791 (2003).

\bibitem{Guo17} W. Guo, X. Liu, Y. Wang, Q. Wei, L. F. Wei, J. Hubmayr, J. Fowler, J. Ullom, L. Vale, M. R. Vissers, and J. Gao, \textit{ Counting near infrared photons with microwave kinetic inductance detectors}, Appl. Phys. Lett. 110, 212601 (2017).

\bibitem{Cheng23}
Cheng, R., Zhou, Y., Wang, S. et al., \textit{A 100-pixel photon-number-resolving detector unveiling photon statistics}, Nat. Photon. 17, 112–119 (2023). https://doi.org/10.1038/s41566-022-01119-3

\bibitem{Pan12} J.W. Pan {\it et al.}, \textit{Multiphoton entanglement and interferometry}, Reviews of Modern Physics {\bf 84}, 777 (2012)


\bibitem{Fabre13} C. Fabre, \textit{Véracité et fausseté en physique. Quelle est la vraie nature de la lumière ? }  Journées Scientifiques annuelles de l'Institut Universitaire de France,Toulouse, April 2-4, 2013.

\bibitem{Werner95} Werner, M. J., Raymer, M. G., Beck, M., Drummond, P. D. (1995). Ultrashort pulsed squeezing by optical parametric amplification. Physical Review A, 52(5), 4202.

\bibitem{Werner97} Werner, M. J.,  Drummond, P. D. (1997). Pulsed quadrature-phase squeezing of solitary waves in $\xi^{(2)}$ parametric waveguides. Physical Review A, 56(2), 1508.

\bibitem{Brambilla04} E. Brambilla, A. Gatti,  M. Bache, and L. A. Lugiato, Simultaneous near-field and far-field spatial quantum correlations in the high-gain regime of parametric down-conversion, Physical Review A {\bf 69} 023802 (2004).


\bibitem{Devaux19} F. Devaux, A. Mosset, and E. Lantz, \textit{Stochastic numerical simulations of a fully spatio-temporal Hong-Ou-Mandel dip}, Phys. Rev. A \textbf{100}, 013845 (2019)

\bibitem{Devaux20} F.Devaux A. Mosset , P. A. Moreau , and  E.Lantz,  \textit{Imaging Spatiotemporal Hong-Ou-Mandel Interference of Biphoton States of Extremely High Schmidt Number}, Phys. Rev. X \textbf{10}, 031031 (2020)





\bibitem{Cahill69}K. E. Cahill and R. J. Glauber, \textit{Density Operators and Quasiprobability Distributions}, Phys. Rev. \textbf{177} 1882 (1969)

\bibitem{Reynaud92}  S. Reynaud, A. Heidmann, E. Giacobino, C. Fabre,\textit{Quantum fluctuations in optical systems}, Progress in Optics {\bf 30}, 3 (1992) 





\bibitem{Mandel95} L. Mandel and E. Wolf, \textit{Optical coherence and quantum optics}, Cambridge University Press (1995) 

\bibitem{Isserli1918}L. Isserlis, \textit{On a formula for the product-moment coefficient of any order of a normal frequency distribution in any number of variables}, Biometrika {\bf 12},  134 (1918)



\bibitem{Trajtenberg2020} S.Trajtenberg-Mills {\it et al.}, \textit{Simulating Correlations of Structured Spontaneously Down-Converted Photon Pairs}, Laser Photonics Rev. \textbf{2020}, 1900321 (2020)


\bibitem{Lantz21} E. Lantz, M. Mabed and F. Devaux, \textit{Violation of Bell inequalities by stochastic simulations of Gaussian States based on their positive Wigner representation}, Physica Scripta {\bf 96}, 045103 (2021)


\bibitem {Kwiatt1995}P.G. Kwiatt, K. Mattle, H. Weinfurter and A. Zeilinger,\textit{ New High-Intensity Source of Polarization-Entangled Photon Pairs}, Phys. Rev. Letters  {\bf 75}, 4337 (1995)

\bibitem{CHSH69} J.F. Clauser, M.A. Horne, A. Shimony and R.A. Holt, \textit{Proposed experiment to test local hidden-variable theories}, Phys. Rev. Lett. {\bf 23},  880 (1969)

\bibitem{Walls08}D.F. Walls and G.J. Milburn, \textit{Quantum Optics}, 2nd  edition, Springer (2008)

\bibitem{Reid86} M. D. Reid and D. F. Valls, \textit{Violations of classical inequalities in quantum optics}, Phys. Rev. A {\bf 34}, 1260 (1986)

\bibitem{Brewster21} R. A. Brewster, G. Baumgartner and Y. K. Chembo,  \textit{Quantum analysis of polarization entanglement degradation induced by multiple-photon-pair generation}, Phys. Rev. A \textbf{104}, 022411 (2021)

\bibitem{Christ11} A. Christ1, K. Lai, A. Eckstein, K. N Cassemiro and C. Silberhorn1, Probing multimode squeezing with correlation functions,New Journal of Physics \textbf{13}, 033027 (2011) 


\bibitem{Massar23} S. Massar, F.Devaux, E.Lantz, \textit{Multiphoton Correlations between Quantum Images}, Phys. Rev. A \textbf{108}, 013705 (2023)

\bibitem{Treps05}N. Treps, C. Fabre, \textit{Criteria of quantum correlation in the measurement of continuous variables in optics}, Laser Physics  {\bf 25}, 187 (2005)

\bibitem{Agafonov10}
I. N. Agafonov, M. V. Chekhova, G. Leuchs, \textit{Two-color bright squeezed vacuum}, Phys. Rev. A {\bf 82}, 011801 (2010)


\bibitem{Jedrkiewicz04}O. Jedrkiewicz {\it et al.}, \textit{Detection of Sub-Shot-Noise Spatial Correlation in High-Gain Parametric Down Conversion}, Phys Rev Letters \textbf{93}, 243601 (2004)

\bibitem{Blanchet08}J.L. Blanchet, F. Devaux, L. Furfaro, and E. Lantz, \textit{Measurement of Sub-Shot-Noise Correlations of Spatial Fluctuations in the Photon-Counting Regime}, Phys. Rev. Letters \textbf{101}, 233604 (2008)

\bibitem{Brida09} G.Brida, L. Caspani, A. Gatti, M.Genovese, A.Meda, I.Ruo-Berchera, \textit{Measurement of sub-shot-noise spatial correlations without subtraction of background}, Phys. Rev. Lett. 102, 213602 (2009) 

\bibitem{OuMandel89} Z.Y Ou, E.C. Gage, B.E. Magill, L. Mandel, \textit{Fourth-order interference technique for determining the coherence length of a light beam}, J.Opt.Soc.Am B {\bf 6}, 100 (1989)

\bibitem{Kim13}Y.S. Kim, O. Slattery, P. S. Kuo, and X. Tang, \textit{Conditions for two-photon interference with coherent pulses}, Phys. Rev. A \textbf{87}, 063843 (2013)



\bibitem{Soro21} G. Soro, E. Lantz, A. Mosset and F. Devaux,
\textit{Quantum spatial correlations imaging through thick scattering media: experiments and comparison with simulations of the biphoton wave function}, Journal of Optics {\bf 23}, 025201 (2021)

\bibitem{Yariv89}A.Yariv, \textit{Quantum electronics, 3d edition}, Wiley, (1989) 

\bibitem{Blanchet10} J. L. Blanchet, F. Devaux, L. Furfaro, and E. Lantz, \textit{Purely spatial coincidences of twin photons in parametric spontaneous down conversion}, Phys.Rev. A \textbf{81}, 043825 (2010)

\bibitem{Eisenberg04}H. S. Eisenberg,G. Khoury,G. A. Durkin,1, C. Simon, D. Bouwmeester, \textit{Quantum Entanglement of a Large Number of Photons}, Phys. Rev. Letters \textbf{93}, 193901 (2004)

\bibitem{Harder16} G. Harder {\it et al.}, \textit{Single-Mode Parametric-Down-Conversion States with 50 Photons as a Source for Mesoscopic Quantum Optics}, Phys. Rev. Letters {\bf 116}, 143601 (2016) 

\bibitem{Mosset04}A. Mosset, F. Devaux, G. Fanjoux, E. Lantz, \textit{Direct experimental characterization of the Bose-Einstein distribution of spatial fluctuations of spontaneous parametric down-conversion}, European Physical Journal D \textbf{28}, 447 (2004) 

\bibitem{Bell87}J.S. Bell,  \textit{Speakable and Unspeakable in Quantum Mechanics}. Cambridge University Press. p. 65 (1987)


\bibitem{Bohm52} Bohm, D. (1952). A suggested interpretation of the quantum theory in terms of" hidden" variables. I. Physical review, 85(2), 166.

\bibitem{Mari12} A. Mari, J. Eisert, \textit{Positive Wigner Functions Render Classical Simulation of Quantum Computation Efficient}, Phys. Rev. Letters \textbf{109}, 230503 (2012)

\bibitem{Rahimi16} S. Rahimi-Keshari, T. C. Ralph, C. M. Caves, \textit{Sufficient Conditions for Efficient Classical Simulation of Quantum Optics}, Phys. Rev. X \textbf{6}, 021039 (2016)


%





\end{thebibliography}
\end{document}